# Capability of Tokai Strainmeter Network to Detect and Locate a Slow Slip: First Results

K. Z. Nanjo [1,2,3,*]

[1] Global Center for Asian and Regional Research, University of Shizuoka, 3-6-1, Takajo, Aoi-ku, Shizuoka, 420-0839, Japan
[2] Center for Integrated Research and Education of Natural Hazards, Shizuoka University, 836, Oya, Suruga-ku, Shizuoka, 422-8529, Japan
[3] The Institute of Statistical Mathematics, 10-3, Midori-cho, Tachikawa, Tokyo, 190-8562, Japan

[*] Correspondence: nanjo@u-shizuoka-ken.ac.jp; Tel.: +81-54-245-5600

ORCID ID https://orcid.org/0000-0003-2867-9185

*Abstract*-The Tokai Strainmeter Network (TSN), a dense network deployed in the Tokai region, which is the easternmost region of the Nankai trough, has been designed to monitor slow slips that reflect changes in the coupling state of the plate boundary. It is important to evaluate the current capability of TSN to detect and locate slow slips. For this purpose, the probability-based magnitude of completeness developed for seismic networks was modified to be applicable to the evaluation of TSN's performance. Using 35 slow slips having moment magnitudes $M$5.1-5.8 recorded by TSN in 2012-2016, this study shows that the probability that TSN detected and located a $M$5-class slow slip is high (> 0.9) when considering a region in and around the TSN. The probability has been found to depend on the slip duration, especially for $M$5.5 or larger, namely the longer the duration, the lower the probability. A possible use of this method to assess the network's performance for cases where virtual stations are added to the existing network was explored. The use of this application when devising a strategic plan of the TSN to extend its coverage westwards is proposed. This extension that allows TSN to cover the entire eastern half of the Nankai trough is important, because the historical records show that the eastern half of this trough tends to rupture first.

*1. Introduction*

The Nankai trough earthquakes occurred with a return period of about 90–200 years. More than 70 years have passed since the last series of Nankai trough earthquakes occurred, the 1944 Tonankai earthquake and the 1946 Nankai earthquake, both belonging to a magnitude 8 class. The possibility of large earthquakes in the Nankai trough has now increased. The earthquakes often occurred in pairs, where a rupture was followed by a rupture elsewhere: for example, the 1854 Ansei-Tokai earthquake and the 1854 Ansei-Nankai earthquake the next day, and the 1944 Tonankai earthquake, followed by the 1946 Nankai earthquake. On the other hand, the fault for the 1707 Hoei earthquake is considered to rupture along its entire length. The history of the Nankai trough earthquakes shows that the eastern half of this trough tends to rupture first (Kanamori 1972; Ando 1975; Ishibashi 2004; Working Group on Disaster Prevention Response when Detecting Anomalous Phenomena along the Nankai Trough 2018). This historical tendency, with current seismic and geodetic observation, implies that the eastern part will rupture earlier than the western part also in the next series of earthquakes, although we cannot say how long the time-lag will be (Nanjo and Yoshida 2018).

The Japan Meteorological Agency (JMA), in collaboration with many institutions and universities, monitors seismicity and crustal deformation of the Nankai trough 24 hours a day (JMA 2017a). Dense networks of instruments accumulate a continuous stream of data related to seismicity, strain, crustal movement, tilt, tidal variations, ground water fluctuations, and other variables. Among them is the strainmeter network in the Tokai region, hereafter referred to as the Tokai Strainmeter Network (TSN), which is operated by JMA (Figure 1). The main task of the TSN is to monitor changes in the coupling state of the plate boundary by detecting and locating slow slips. Thus, the reliable evaluation of TSN's monitoring ability of slow slips is important.

Slow slips that grow over time may result in large earthquakes, as was observed for the 2011 Tohoku-oki earthquake of moment magnitude $M9$ (Kato et al. 2012). As another example of a precursor phenomenon just before an earthquake occurs, tilting associated with the 1944 $M8$-class Tonankai earthquake has been observed (Mogi 1984). A remarkable precursory tilt started two or three days before the earthquake. The precursory amount of change corresponds to about 30% of the amount of change at the time of the earthquake. In contrast, many studies using strainmeters and tiltmeters located close to the eventual earthquakes have concluded that a precursory slip, if any, is very small, $< 1\%$, for many California earthquakes (Wyatt 1988; Kanamori 1996) such as the 1987 Whittier Narrows earthquake ($M = 6.0$), the 1987 Superstition Hills earthquake ($M = 6.6$), the 1989 Loma Prieta earthquake ($M = 6.9$), and the 1992 Landers earthquake ($M = 7.3$).

Scholz et al. (1972) made detailed laboratory measurements on frictional characteristics of granite. In the condition where the stick-slip predominated, the stick-slip was preceded by a small amount of stable slip, which accounted for about 2-5% of the unstable slip that followed. Lorenzetti and Tullis (1989) used mechanical models of faulting during an earthquake cycle that was based on the rate- and state-dependent friction law, and predicted that the amount of pre-seismic moment release was $< 0.5\%$ of the earthquake moment for most simulations. Using a simulation of the earthquake cycle in the Tokai region, Kato and Hirasawa (1996) did not directly estimate $M$ of a precursory slow slip. However, their result indicates that $M$ varies, depending on values given for different parameters. Furthermore, in their earthquake-cycle simulation, it was commonly seen that remarkable abnormal crustal movements started to appear over a wide area several days to several hours before the occurrence of an earthquake.

It is difficult at present to determine $M$ of slow slips that are precursory phenomena to the Nankai trough earthquake, but it is necessary to assume a severe situation. Based on the discussion provided above, a precursor slip would be assumed to be $< 0.5\sim1\%$ of the earthquake. As the Nankai trough earthquakes belong to a $M8$-class or larger, the final precursory slow slip with $< M6\sim6.5$ is assumed. Under this assumption, it could be of high value to detect and locate slow slips when they are still small. In this study, we estimated the capability of TSN to detect and locate a slow slip.

The approach was based on the Probabilistic Magnitude of Completeness (PMC) method, which had been developed for seismic networks (e.g., Bachmann et al. 2005; Schorlemmer and Woessner 2008; Nanjo et al 2010; Schorlemmer et al. 2018). This study is the first to be applied to a strainmeter network. This PMC-based method relies on empirical data. Station probability ($P_{st}$), which is the probability that a station was used to detect and locate a slow slip of a given magnitude $M$ at a given distance $L$, where $L$ is distance in the three spatial dimensions, is derived. From the $P_{st}$ for all stations, we obtain the probability ($P_{dl}$) of detecting and locating slow slips of $M$ at a point ($x$) for three or more stations. If this method is applied to TSN, the product is a map of $P_{dl}$ for $M$. A feedback for JMA network operators is foreseen by providing a tool to infer spatial heterogeneity of current monitoring ability. This tool will be used as a basis to help future planning for optimizing network coverage.

JMA, with a similar motivation as that shown in this study, mapped the lower-limit $M$ ($M_{LL}$) to detect and locate slow slips in the Nankai trough (Research Committee for Forecast Ability of Large Earthquakes along the Nankai Trough 2017). This mapping calculated crustal deformation due to a slow slip of given $M$ at a point $x$ on the plate interface using the method of internal deformation due to shear fault in an elastic half-space (Okada 1992). The strike and dip of a fault at $x$ were determined in such a way that the fault is a tangent plane to the plate boundary interface at $x$ (Kobayashi 2000). The slip angle of the fault was based on the direction of plate convergence (Heki and Miyazaki 2001). If crustal deformation due to a slow slip of $M$ at $x$ causes strain changes that are satisfied by a criterion that the signal-to-noise ratio is larger than a predefined value at three or more stations, then this slow slip is considered to be detected and located. For each $x$, the $M_{LL}$ to be satisfied with the criterion is searched, and this is mapped at $x$. However, it is nontrivial to assign noise levels for each station because tidal correction (Ishiguro et al. 1984; Tamura et al. 1991), geomagnetic correction (multicomponent strainmeters) (Suganuma et al. 2005; Miyaoka 2011), and precipitation and atmospheric-pressure correction (volumetric strainmeters) (Hikawa et al. 1983; Ishigaki 1995; Gebauer et al. 2010) are applied as pre-processes for each analysis, where the geomagnetic correction is needed, because multicomponent strainmeters use magnetic sensor to measure radical deformation (volumetric ones do not use this sensor), so that the output is affected by magnetic fields. These noise levels for each station vary with time (Miyaoka and Yokota 2012). Moreover, setting a criterion of signal-to-noise ratio, for example, 2 or 3 as was done in Miyaoka and Yokota (2012), is arbitrary when calculating $M_{LL}$.

The significance of this research is that a criterion was not predefined for the signal-to-noise ratio, nor was the noise level assigned to each station, even though both are critical for the computation of $M_{LL}$. In other words, the $P_{dl}$ computation requires no assumption once $P_{st}$ is determined based on empirical data. Thus, mapping $P_{dl}$ provides an alternative to mapping $M_{LL}$. This adds value to the current state of the art of reliable estimation of detection-location capability of a slow slip because changes in the coupling state of the plate boundary will not be missed even if they are small from the viewpoint of disaster prevention measures. The results obtained in this study show general agreement with JMA's results, although both methods are based on different ideas and approaches, which are outlined in the "Discussion" section.

## 2. Methods

The PMC-based method relies on two sources of data: (1) station data describing the location for each station in the network; (2) the slow-slip catalogue describing the location, time, and $M$ for each slow slip including data describing which stations were used to detect and locate this slow slip. The method is divided into an analysis part and a synthesis part. See (e.g., Bachmann et al. 2005; Schorlemmer and Woessner 2008; Nanjo et al 2010; Schorlemmer et al. 2018) for applications to seismic networks.

### 2.1. Analysis part

In the analysis, data triplets were first compiled for each station. A triplet contains, for each slow slip, (i) information on whether or not this station was used for detecting and locating the event, (ii) the $M$ of the event, and (iii) its distance $L$ from the station. Figure 2 provides an illustrative example to show how to generate data triplets. Assume that a strainmeter network consisting of more than 3 stations has detected and located 5 slow slips, where the $i$-th event is called event $i$ ($i = 1, 2, …, 5$). The left panel of Figure 2 shows that station 1 was used to detect and locate event 1, event 2, and event 4 (green), but not events 3 and 5 (red). Note that events 3 and 5 were recorded by using other stations. The data triplet of event 1 contains (i) the fact that station 1 was used to detect and locate this event (green), (ii) the moment magnitude ($M_1$), and (iii) its distance from station 1 ($L_1$). If a station was used to detected and locate an event, the data triplet of this event is referred to as the "plus triplet" for the station, otherwise as the "minus triplet". For station 1, data triplets of events 1, 2, and 4 are plus triplets, and those of events 3 and 5 are minus triplets. These triplets are plotted in the graph of $L$ as a function of $M$. The center panels of Figure 2 show plots of data triplets of stations 2 and 3. The same applies for all stations.

Using triplets for each station, the desire was to determine $P_{st}(M, L)$, the probability that the station detects changes in strain associated with a slow slip for a given set of $M$ and $L$ to locate the event. As described below, the number of slow slips used for the analysis is 35, which is not enough to reliably estimate $P_{st}(M, L)$ for individual stations. For a more reliable estimate of $P_{st}(M, L)$, we used the idea of Bachmann et al. (2005) in which data triplets from all stations that have been used at least once for the period of interest are stacked (Figure 2).

Figure 3a shows data triplets for all stations stacked and plotted in the *L-M* graph. Using data triplets close to a given pair $(M, L)$, $P_{st}(M, L)$ was computed by the number of plus triplets, $N_+$ (green plus symbol), divided by the sum of $N_+$ and the number of minus triplets, $N_-$ (red minus symbol): $P_{st}(M, L) = N_+/(N_+ + N_-)$ (Figure 3b) (e.g., Schorlemmer and Woessner 2008; Nanjo et al 2010; Schorlemmer et al. 2018).

$P_{st}(M, L)$ was smoothed by applying a simple constraint: $P_{st}$ cannot decrease with smaller *L* for the same *M* (e.g., Schorlemmer and Woessner 2008; Nanjo et al 2010; Schorlemmer et al. 2018). This smoothing accounts for high probabilities at short distances (Figure 3b). Another constraint, namely that the smoothed probability cannot increase with decreasing *M* at the same distance, was not applied because imposing this constraint would lead to an overestimation of $P_{st}(M, L)$.

One problem arises for stations that have never been used for event detection and location even though these stations were in operation. They were relatively far from the slow slips during the observed period (2012-2016), assuming that JMA only uses these stations if slow slips occur near them. Consequently, $P_{st}(M, L)$ was assigned for all stations in operation, regardless of whether or not they have been used for event detection and location.

### 2.2. Synthesis part

In the synthesis part, basic combinatorics were used to obtain $P_{dl}(M, x)$ for detecting and locating a slow slip of *M* at location *x*, given a specific network configuration (e.g., Bachmann et al. 2005; Schorlemmer and Woessner 2008; Nanjo et al 2010; Schorlemmer et al. 2018). $P_{dl}(M, x)$ for TSN is defined as the probability that three or more stations detect changes in a strain associated with a slow slip of *M* to locate the event at *x*. Only two are needed if admitting that a slow slip occurs on the plate boundary interface, where the depth contours are shown by red curves in Figure 1. However, JMA requires at least three stations (Research Committee for Forecast Ability of Large Earthquakes along the Nankai Trough 2017; Miyaoka and Yokota 2012). The minimum number of stations must be adjusted if the condition of the TSN is based on another number of stations.

If $P_{dl}$ is larger than a threshold, this is considered as an indication that a slow slip will not be missed. Previous researchers (Schorlemmer and Woessner 2008; Nanjo et al 2010; Schorlemmer et al. 2018) took several values among 0.99~0.99999 as a threshold. Given the considerable uncertainty in all phenomena such as tectonic strain accumulation, slow slips, and others in this study, the threshold needs to be as high as possible to secure that a slow slip will not be missed. We assumed a conservative value of $P_{dl} = 0.9999$, where the complementary probability $Q (= 1 - P_{dl})$ that a slow slip will be missed is 0.0001 $(= 1 - 0.9999)$. We avoided smaller *Q* because of possible computational artifacts: our preliminary study experienced that the solution often did not converge when taking $P_{dl} = 0.99999$ as the threshold.

### 3. Data

JMA is recording short-term slow slips in the Tokai region with the TSN, one of the densest networks in Japan, operating 11 multicomponent strainmeter stations and 16 volumetric strainmeter stations (Figure 1). The former strainmeters are generally buried in boreholes at depths of 400-800 m, while the latter ones are in the range of 150-250 m. The station list can be obtained from JMA. Note that for the time being, there is no tiltmeter included in the TSN (JMA 2017b).

Short-term slow slips in the Tokai region release energy over a period of a few days to a week, rather than seconds to minutes which is characteristic of a typical earthquake. Long-term slow slips that slip over a period of a few months to several years were recorded by GNSS (Global Navigation Satellite System). In the Tokai region, long-term slow slips of *M*7.0 and *M*6.8 occurred in 2001-2005 and 2013-2017, respectively. Short-term slow slips can be seen at depths of 30-40 km, and the down-dip side of the long-term slow slips is at a depth of 20-30 km. There is interest in knowing the capability of TSN to detect and locate small slow slips. This study used the short-term slow-slip catalogue (Kimura and Miyaoka 2017), which includes 35 events for *M*5.1-5.8 with depths 26-41 km from 2012 to 2016 in the Tokai region (Figure 1). Based on the slow-slip catalog and the list of strainmeter stations, data triplets were compiled.

It would be valuable for the reader to see one event in a figure exemplarily as short timeseries. Figure 4a shows changes in strain associated with a slow slip of *M*5.5 that occurred during 24-25 August 2015 (duration is 2 days), located at the point (circle in Figure 4b), observed by the four multicomponent strainmeter stations (filled triangles in Figure 4b). The waveform shown here is the one after the physics-based tidal correction using Baytap-G (Ishiguro et al. 1984; Tamura et al. 1991), and the statistics-based geomagnetic correction (Suganuma et al. 2005; Miyaoka 2011) were applied as pre-processes. Because the strainmeters in this case are multicomponent, corrections for volumetric strainmeters, such as physics-based precipitation correction using the autoregressive (AR) model (Ishigaki 1995) and statistics-based atmospheric-pressure correction (Hikawa et al. 1983), are not applied. The onset time and ending time of this slow slip were determined by seeing these waveform traces with the naked eye of the JMA network operators (24-25 August, shown by light blue in Figure 4a). The location and magnitude of this slow slip were those that best explained the observed changes in strain shown in Figure 4a, using the method of internal deformation due to shear fault (Okada 1992; Nakamura and Takenaka 2004), where the slow slip was resolved at the plate boundary interface (Figure 4b). Uncertainty in $M$ is 0.3 in this case, while it is typically 0.1-0.3. To overcome the situation that identifying a slow slip only from waveform in strain is usually difficult, the operators use the information on low-frequency earthquakes (LFEs) that often occur with slow slips. Detecting and locating LFEs by the seismic networks to create the JMA earthquake catalog, a different catalog from the slow-slip catalog, trigger the operators to review waveform traces for stations near the epicenters of LFEs.

The method uses the distance between the source and the stations, and a measure of the size of the source (moment). Although rupture velocity does not vary much from one earthquake to another, this is no longer true for slow slips. The duration may then be of importance, and the same total slip (hence moment) over longer durations will imply that $P_{st}(M, L)$ should be lower. In a preliminary analysis, $M$ was plotted as a function of duration (Figure 5). Duration was defined by the time difference between the starting time of a slip and the ending time of the slip. $M$ varied from 5.1 to 5.8 for short durations ($\leq 5$ days), while for longer durations ($> 5$ days), the range of $M$ was 5.4 to 5.7. The duration dependence on $P_{st}(M, L)$ and $P_{dl}(M, x)$ is verified in section 4.3.

To compute $P_{st}(M, L)$, data triplets were used, each having $M$ and $L$ close to a given pair $(M, L)$. Triplets were selected by measuring the distance between each triplet and pair $(M, L)$. To measure such a distance, a metric in the $M$-$L$ space needed to be defined. Schorlemmer and Woessner (2008) proposed the use of an attenuation equation for magnitude-based determination of earthquakes located in a given local seismic network, and redefined a metric in the transformed magnitude-magnitude space. In this study, we followed this idea and used the attenuation equation used by JMA (Tsuboi 1954; Schorlemmer et al. 2018). The magnitude of a slow slip is defined by the moment magnitude ($M$), and not by the JMA magnitude ($M_{JMA}$). However, when $M$ exceeds 5, $M_{JMA}$ can be considered statistically equivalent to $M$, while below $M = 5$, $M_{JMA}$ is smaller than $M$ (JMA 2003; Scordilis 2005). The slow slips that were analyzed in this study had $M > 5$. Thus, it was assumed, for our case, that the attenuation equation used by JMA is directly applicable to define a metric in the transformed magnitude-magnitude space. Given this metric, we selected all triplets that obey the criterion of a metric smaller than or equal to 0.2, which is a usual magnitude error.

This approach assumes that single slow slips occur at different times because multiple slow slips that had occurred at different locations at the same time were not reported by JMA during 2012-2016.

Figures 3a,b show the distribution of stacked data triplets and the corresponding distribution of $P_{st}(M, L)$, respectively. There is no triplet for short distances ($L < 35$ km) due to the distances from stations to the plate interface on which slow slips occurred. Despite this, $P_{st} > 0.9$ (dark green) is seen, irrespective of $M$. This is because the smoothing constraint described in section 2.1 was applied. General patterns of $P_{st}$ for $M = 5.1$-5.4 are similar to each other: there is a band of $P_{st} \sim 0.8$ (yellow) around $L = 40$ km, above which $P_{st}$ decreases with $L$ (red). It was observed that the patterns for $M = 5.5$-5.8 are different from those for $M = 5.1$-5.4: Events of higher $M$ can be observed at greater distances, thus $P_{st}$ is higher, while events of lower $M$ are observable only at shorter distances and thus have lower $P_{st}$ at long distances $L > 40$ km. Before applying $P_{st}$ to create the maps of $P_{dl}$, simple sensitivity checks on dependence of volumetric and multicomponent strainmeters were conducted on the distribution of data triplets, which are shown in the next section.

## 4. Results

### 4.1. Sensitivity checks on the dependence of volumetric and multicomponent strainmeters affect distribution of data triplets

We first performed simple sensitivity checks on dependence of four different azimuths of multicomponent strainmeters of a station on the distribution of triplets. The strainmeters adopt magnetic sensors to measure radial deformation (contraction and extension) of the cylindrical vessel in four directions separated by nearly 45 degrees. Although data were sparse, the distribution of triplets for each of the four azimuths were separately considered. Figure 6 shows an example from station "Tahara Takamatsu". Generally, the patterns are similar to each other. This is due to the fact that for most cases, strain changes for all four components were used to detect and locate slow slips. The same trend was observed for the other multicomponent stations. Thus, for each multicomponent strainmeter station, we stacked triplets from the four components to define these as data triplets of that station.

A similar sensitivity analysis was performed for dependence of volumetric and multicomponent strainmeters on the distribution of triplets. "Tahara Fukue" (volumetric strainmeter station) and "Tahara Takamatsu" (multicomponent strainmeter station), which are spatially close to each other (Figure 7), were selected. The former measures volumetric change while the latter measures changes in diameter (line strain) of the four azimuths. Generally, the patterns were quite similar to each other. We treated all stations equally regardless of whether strainmeters were volumetric or multicomponent in nature.

### 4.2. Mapping $P_{dl}$ in and around TSN: current capability of TSN to detect and locate a slow slip

Figure 8 shows the spatial distribution of $P_{dl}$ for $M = 5.1$, 5.3, 5.5, and 5.8 (from the minimum $M$ to maximum $M$ of observed slow slips), where see also Figure 1 of Online Resource showing maps for $M = 5.1$, 5.2, …, and 5.8. We used a grid spacing of $0.05° \times 0.05°$ and computed $P_{dl}$ at the place boundary interface (Baba et al. 2002; Nakajima and Hasegawa 2007; Hirose et al. 2008; Nakajima et al. 2009). The four stations in the eastern part of TSN on the peninsula were really used and included in the probability calculations. Because the probability maps in Figure 8 were resolved on the plate boundary interface, this peninsula, below which there is no plate boundary interface, is not colored. Among all stations in TSN, three showed a station characteristic in which the occurrence frequency of irregular changes in the strain was high (cyan triangles in Figure 8) (Miyaoka and Yokota 2012). Note that they are in operation but have never been used for event detection and location. These three stations were not included into the $P_{dl}$ computation. Figure 8 shows that slow slips during 2012-2016 are located in areas with predominantly high probabilities, supporting a consistency between our synthesis and the observation.

We confirmed the expectation that a region of high probabilities ($P_{dl} \geq 0.9$) for $M = 5.1$-5.8 almost covers the TSN. Detailed characteristics are as follows. As expected from $P_{st}$ in Figure 3, the spatial patterns of $P_{dl}$ for $M = 5.1$ and 5.3 are similar to each other (Figure 8a,b). The pattern of very high probabilities ($P_{dl} \geq 0.9999$) is spatially heterogeneous: a noticeable feature is that $P_{dl} \geq 0.9999$ is not seen at the near-coast offshore around 137.5°E but around other longitudes. This may be due to the lack of stations around 137.5°E near the coast. Above $M = 5.4$ (Figure 8c,d), the region of $P_{dl} \geq 0.9$ increased with $M$. The anticipated source zone (grey chain line) of the Tokai earthquake, the easternmost segment in the Nankai trough, is covered by a region of $P_{dl} \geq 0.9$ for $M = 5.8$ (Figure 8d). When considering $P_{dl} \geq 0.9999$ as an indication that a slow slip will not be missed, a slow slip with $M \leq 5.8$ would likely be missed in a southern part of the anticipated source zone.

### 4.3. Dependence of duration of a slow-slip on $P_{st}$ and $P_{dl}$: the longer the duration, the greater the difficulty in detecting and locating a slow slip

To examine whether the duration of a slow slip influences $P_{st}$ and $P_{dl}$, we divided all of the data triplets into two datasets: one with a duration of $\leq 5$ days and the other with a duration of $> 5$ days. For each dataset, we computed $P_{st}$ and created $P_{dl}$ maps in Figures 9 and 10 (see also Figures 2 and 3 of Online Resource), where computing and mapping procedures are the same as those for Figures 3b and 8, respectively. Since there was no slow slip of $M < 5.4$ for a duration $> 5$ days (Figure 5), $P_{st} = 0$ was observed for $M < 5.4$ in Figure 10c. $P_{st}(M < 5.4, L)$ for a duration of $\leq 5$ days (Figure 9e) is the same as $P_{st}(M < 5.4, L)$ in Figure 3b, resulting in the same maps of $P_{dl}(M = 5.1$ and 5.3, $x$) in Figures 8a,b and Figures 9a,b.

For a short duration (Figure 9e), $P_{st}$ at large distances (e.g., $L > 40$ km) increased with $M$, but it did not increase for a long duration (Figure 10c). The difference in $P_{st}$ between short duration (Figure 9e) and long duration (Figure 10c) is clear for $M = 5.6$-5.8. As expected from $P_{st}$, $P_{dl}(M = 5.5, x)$ in Figures 9c is similar to that in Figures 10a. A comparison between Figures 9d and Figures 10b for $M = 5.8$ shows that the capability of detecting and locating a slow slip is lower for a long duration ($> 5$ days) than for a short duration ($\leq 5$ days). Changes in strain produced by a long-duration slow slip were more gradual than that by a short-duration slow slip. The former changes were more difficult to be distinguished from background levels in strain than the latter changes. This is plausibly the reason why the detection and location of a slow slip is more critical for a long duration than for a short one.

### 4.4. Virtual installation of one or more stations into TSN

To infer the effect of adding station(s) to the TSN on $P_{dl}$, scenario computations were performed by virtually placing additional stations to the network configuration. A fundamental problem is the definition of $P_{st}$ that is used for individual stations installed for the virtual case. As was applied for stations that have never been used for event detection and location even though these stations were in operation, $P_{st}(M, L)$ was assigned for virtual stations.

Figure 11 shows the maps of $P_{dl}$ for $M = 5.5$ and 5.8 for virtual station installations at different locations in addition to the existing network. $P_{dl}$ was considered based on $P_{st}$ obtained for all slow slips, irrespective of their durations. A single virtual station was added at a location (Figures 11a,c). At this location, one volumetric strainmeter station was operating. However, as described in section 4.2, this station was not included into the $P_{dl}$ computation for creating the maps in Figures 8-10, because a characteristic of the station was the frequent occurrence of irregular changes in the strain (Miyaoka and Yokota 2012). Replacement with a new single station without such station characteristic was assumed and $P_{dl}$ was computed for this virtual network configuration (Figures 11a,c). Comparison with Figures 8c,d shows that this replacement improved the detection-location capabilities, especially increasing $P_{dl}$ for $M = 5.5$ in an offshore region near the coast around 137.5°E. Increasing the number of stations at locations close to each other (2 more stations in Figure 11b,d), which would further enhance the capabilities in the same region, was considered next.

*5. Discussion*

As pointed out in the "Introduction" section, it is a standard practice to make severe assumption that a precursor slip would be $< 0.5~1\%$ of the earthquake. The final precursory slow slip with $< M6~6.5$ to the Nankai trough earthquakes belong to a $M8$-class or larger is assumed, if it exists. This assumption implies that it is not necessary for precursory slips to have an observable size (e.g., the occurrence of slips of $M4$-class or smaller is a possibility). Furthermore, the current study used only 35 slow slips and presented the first attempt of applying the PMC-based method to the evaluation of TSN's performance. Thus, readers should be careful about extrapolating these results to precursory slips of a megathrust earthquake in the Nankai trough. However, if the primary purpose of the TSN is considered, it is important to monitor changes in the coupling state of the plate boundary by detecting and locating slow slips. The estimation made in this study of the capability of TSN to detect and locate a slow slip provides the first results. Future research would use much more data to constrain $P_{dl}$ with sufficient certainty.

The early detection-location capability of a slow slip has not yet been considered. Data obtained by strainmeters under TSN are being monitored continuously at JMA, so that rapid earthquake information is in operation in real-time (JMA 2017a). Miyaoka and others developed a stacking method in which data at different strainmeter stations are added to increase the signal-to-noise ratio for early detection of crustal deformation associated with slow slips (Miyaoka and Yokota 2012; Miyaoka and Kimura 2016). This method, in combination with the PMC-based approach reported in this study, will lead to a more realistic evaluation of TSN's performance regarding the early detection-location capability of a slow slip.

A conventional assessment of $M_{LL}$ has been applied to the entire Nankai trough (Research Committee for Forecast Ability of Large Earthquakes along the Nankai Trough 2017) by using a model assumption that the medium is elastic and that the strainmeters record elastic strain changes caused by slow slips (Okada 1992). This paper addressed a fundamental question, namely whether this PMC-based method and the conventional method gave similar results. The map of $M_{LL}$ for the Nankai trough was created based on the TSN operated by JMA and the network operated by AIST (National Institute of Advanced Industrial Science and Technology). Since only three AIST stations are in operation in the Tokai region, it was assumed that the spatial pattern of $M_{LL}$ based on this hybrid network was comparable to the maps of $P_{dl}$ purely based on TSN. Values of $M_{LL} \leq 5.8$ fall in and around the hybrid network in the Tokai region, where slip duration was not taken into consideration for the $M_{LL}$ computation. Given that $P_{dl} \geq 0.9999$ is an indication that a slow slip will not be missed, the probability map of $M = 5.8$ (Figure 8d), where values of $P_{dl} \geq 0.9999$ fall in and around TSN, shows consistency with the $M_{LL}$ map. When a small magnitude ($M = 5.5$) was assumed, this again demonstrated general agreement between the region of $P_{dl} \geq 0.9999$ (Figure 8c) and the region of $M_{LL} \leq 5.5$, although detailed differences in patterns demonstrate that the former region spreads wider toward offshore at 138.0-138.5 °E than the latter region. Regardless of the different approach used in this study relative to a conventional approach, the results were generally similar to each other. However, these constitute the first results that need to be supported by much more data in future.

Monitoring changes in the coupling state of the Nankai trough plate boundary may provide researchers with qualitative information on the increased (or decreased) possibility of the occurrence of an impending large earthquake (JMA 2017a). It is important to estimate the coupling state through the occurrence of slow slips. Moreover, for such an estimation, it is vital to reliably evaluate monitoring ability of slow slips, as was shown in this study and in a previous study (Research Committee for Forecast Ability of Large Earthquakes along the Nankai Trough 2017). However, forecasting $M$ and timing of an earthquake involves large uncertainty from a scientific viewpoint. For example, if slow slips have a duration of 5 days, could there be knowledge about "precursor" to the eventual great earthquake that might occur after these 5 days or later? In other words, what is the time span until the eventual one would happen? Based on the present research, we cannot say what time span is left until the earthquake starts, implying no knowledge about time to evacuate people from the future Nankai trough earthquake. Thus, considering that earthquakes occur suddenly is a major premise to implement disaster prevention measures, but the remaining damage can be huge even if a response occurs. When abnormal phenomena related to plate coupling are observed, it is necessary to make full use of such information for reducing disasters. For example, it is known, from a study of a scenario of Nankai trough earthquakes, that the predicted time elapse until a 1-meter tsunami wave arrives, is a few minutes to several tens of minutes (Shizuoka Prefecture 2014). If the possibility of the occurrence of an impending large earthquake is judged to increase, this information may be used as a trigger to evacuate, in advance, elderly people who live near the coast to a safe place where they are expected to stay for a certain period (Working Group on Disaster Prevention Response when Detecting Anomalous Phenomena along the Nankai Trough 2018).

## *6. Conclusion*

The current capability of TSN to detect and locate a slow slip was evaluated in this study. The PMC method for seismic networks was modified to be applicable to the evaluation of TSN's performance. A currently available catalog, in which 35 slow slips with $M = 5.1$-$5.8$ (depth of 26-41 km) in 2012-2016, recorded by TSN, was used. It was confirmed that a region of high probabilities ($P_{dl} \geq 0.9$) that TSN detected and located a slow slip of $M = 5.1$-$5.8$ almost covered the TSN. In more detail, the spatial patterns of $P_{dl}$ for $M = 5.1$-$5.4$ were similar to each other (Figures 8a,b). Above $M = 5.4$ (Figures 8c,d), the region of high $P_{dl}$ values ($P_{dl} \geq 0.9$) increased with $M$. If $P_{dl} \geq 0.9999$ is interpreted as an indication that a slow slip will not be missed, a slow slip with $M \leq 5.8$ would likely be missed in a southern part of the anticipated source zone of a future Tokai earthquake.

It was further shown in Figures 9 and 10 that $P_{dl}$ is generally lower for slips of long duration (> 5 days) than for those of short duration (≤ 5 days), where the duration was defined by subtracting the starting time of a slip from the ending time of the slip. This result implies that the longer the duration, the greater the difficulty in detecting and locating a slow slip. Gradual changes in strain produced by a long-duration slow slip were difficult to be distinguished from background levels, compared with the changes in strain by a short-duration slow slip. This is a physically plausible reason of the findings that possibilities of detecting and locating a slow slip event are higher when the duration of the event is shorter. The $P_{dl}$ maps created by using slow slips regardless of their duration (Figure 8) are considered to show the average capability of TSN to detect and locate slow slips over short- and long-durations.

Given the reported rupture history (Kanamori 1972; Ando 1975; Ishibashi 2004; Working Group on Disaster Prevention Response when Detecting Anomalous Phenomena along the Nankai Trough 2018, Nanjo and Yoshida 2018) described in the "Introduction" section, it is desirable to explore the possibility of making a strategic plan for TSN to extend its coverage westwards to the entire eastern half of the Nankai trough. Note that for the entire eastern half of the Nankai trough, the detection-location capability of a slow slip is currently low, except for the Tokai region and a part of the Kii Peninsula (Research Committee for Forecast Ability of Large Earthquakes along the Nankai Trough 2017). For this purpose, a tool proposed in this study can help on network planning with simulation of virtual station installation (Figure 11). However, it is understood that the effectiveness of this tool needs to be investigated in more detail as it is a non-trivial task to assume a station characteristic for a new station. Additional information such as local site conditions and geological parameters need to be available. Nonetheless, as a rule of thumb, the PMC-based method can help, with reduced costs, to estimate network performance and infer locations for future stations.

Cases where additional submarine stations are virtually placed in offshore regions were not considered because it was assumed that $P_{st}$ for a virtual seafloor station was not the same as that used for an inland station. As seen in section 4.4, virtual installation of one or more inland stations would certainly increase $P_{dl}$ in far-offshore regions, but the effect is limited. Two seafloor strainmeter stations under DONET (Dense

Oceanfloor Network system for Earthquakes and Tsunamis), not involved in TSN, are operating in a far-offshore region from the Kii peninsula near the trough axis (33.0-33.5°N, 136.0-136.5°E) (Araki et al 2017). Eight slow slips in 2011-2016 have been recorded thus far. The next generation for evaluating TSN's performance may make use of information of seafloor strainmeter records.

## *Acknowledgments*

The author thanks two anonymous reviewers for their useful comments. The author would also like to acknowledge H. Kimura and K. Miyaoka for providing slow-slip and strainmeter-station catalogues (Kimura and Miyaoka 2017), and Y. Ishikawa and J. Kasahara for discussions. Strainmeter stations in TSN were installed by JMA, except for the stations "Hamamatsu Haruno" and "Kawanehonshou Higashifujikawa" that were installed by Shizuoka Prefecture. This work was partially supported by JSPS KAKENHI Grant Number JP 17K18958 and the Ministry of Education, Culture, Sports, Science and Technology (MEXT) of Japan, under its Earthquake and Volcano Hazards Observation and Research Program. Some figures were produced by using GMT software (Wessel et al. 2013). Data are available upon reasonable request.

## REREFENCES

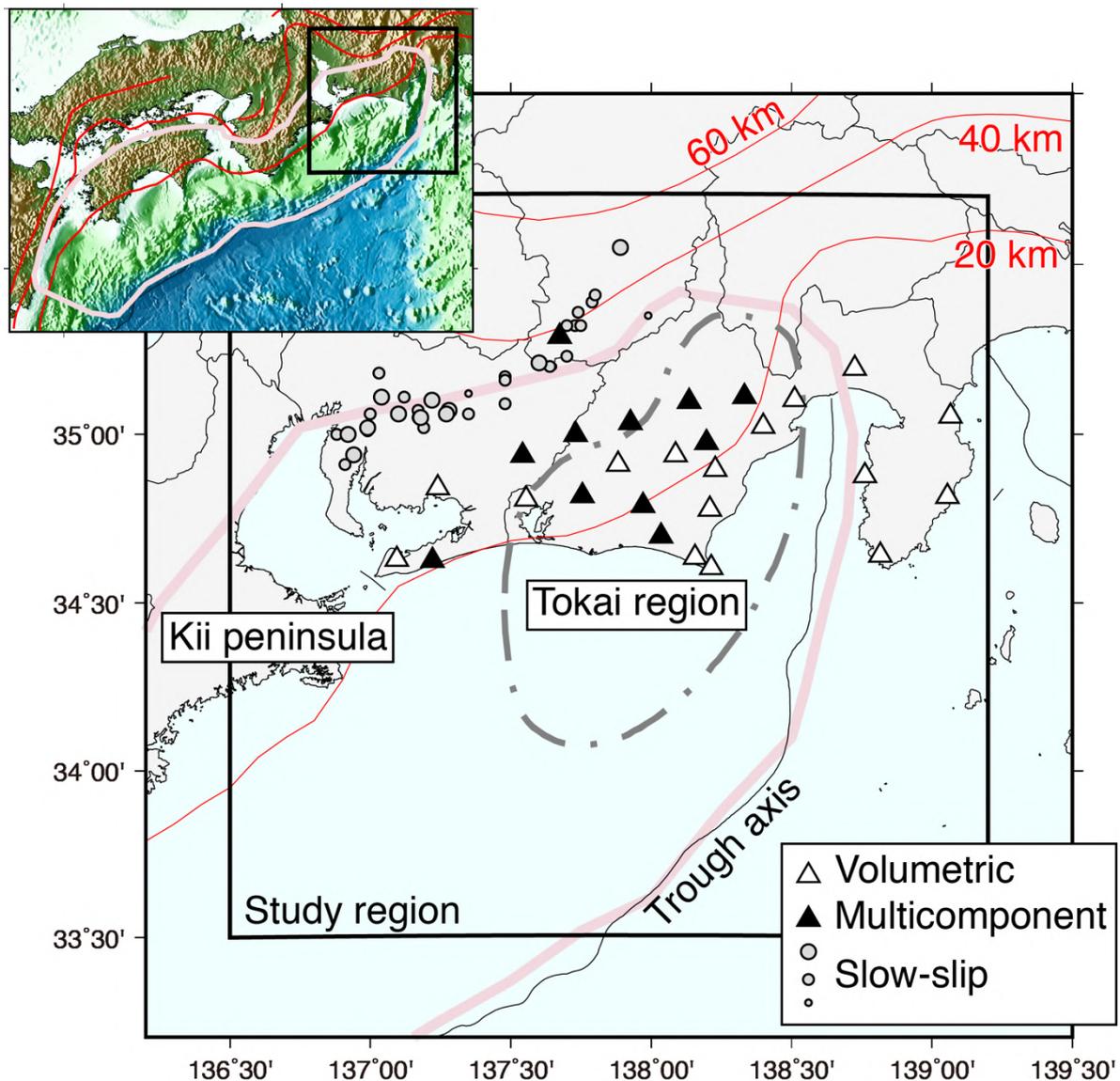

**Fig. 1** Location map of strainmeter stations (open triangle: volumetric, filled triangle: multicomponent) in the Tokai region. Slow slip (circle): small size, $M \geq 5.1$; middle size, $M \geq 5.4$; large size, $M \geq 5.7$. Chain line: region supposed to be the source of an anticipated Tokai earthquake (https://www.data.jma.go.jp/svd/eqev/data/nteq/tokaieq.html). Thick purple line: maximum focal region of a megathrust earthquake (http://www.bousai.go.jp/jishin/nankai/model/pdf/chukan_point.pdf). Red curves mark depth contour lines of the plate boundary interface (Baba et al. 2002; Nakajima and Hasegawa 2007; Hirose et al. 2008; Nakajima et al. 2009). Prefectural boundaries are shown by thin black lines. The zoomed-out inset is a map of the Nankai trough, where the study area (black square) is shown.

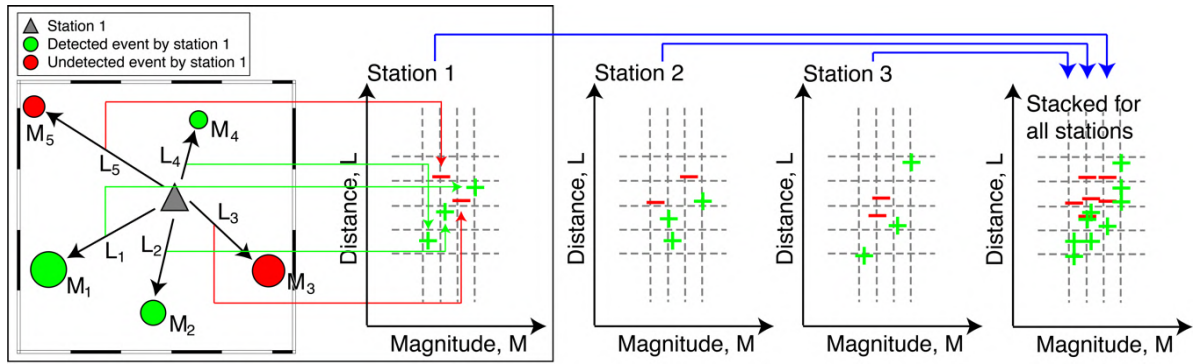

**Fig. 2** Illustration showing the procedure to create a stacked distribution of detected slow slip events and undetected events. Assume that 5 events (event 1, event 2, …, event 5) are detected and located by a strainmeter network, where the moment magnitude of an event $i$ is $M_i$ with event number $i = 1, 2, …, 5$. The left panel shows a spatial map of these events relative to strainmeter station 1 ($L_i$: distance of event $i$ from station 1) and a graph of $L$ as a function of $M$. If station 1 was used to detect and locate a slow slip, the data triplet of this event is called a "plus triplet", colored in green in the map and plotted in the graph by a green plus symbol. If station 1 was not used, the data triplet is called a "minus triplet" (red, minus symbol). Central panels of stations 2 and 3 show the corresponding $L$-$M$ graphs. For the sake of brevity, the graphs are shown for three stations. However, we assume more than three stations operating under the network. The right panel shows the graph of $L$ versus $M$, in which data triplets stacked for all stations are plotted. In this illustrative manner, Figure 3a was created.

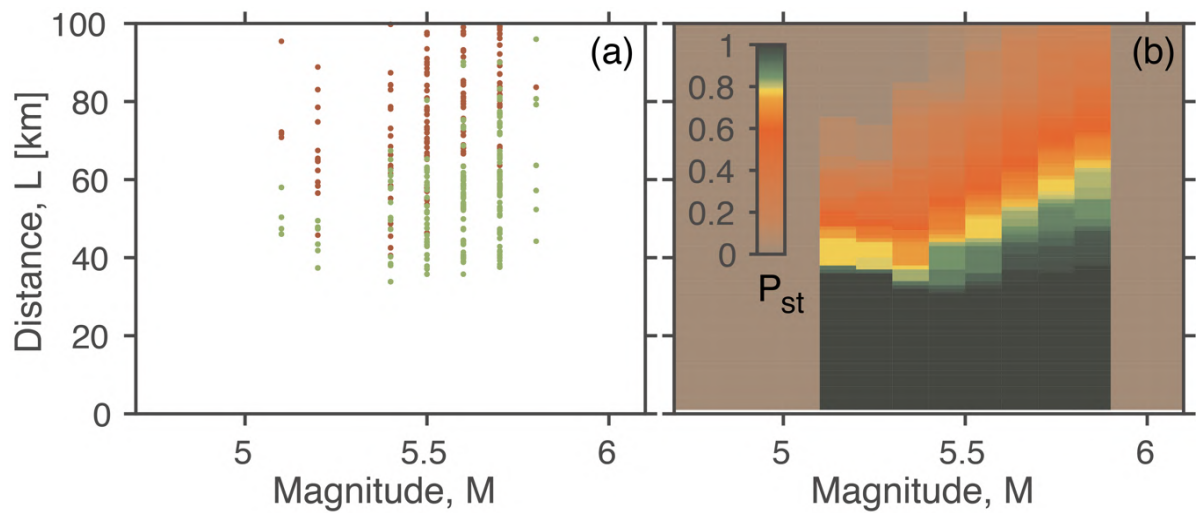

**Fig. 3** Average station characteristics. (a) Stacked distribution of undetected events (red) and detected events (green) for stations that have been used at least once to record slow slips in 2012-2016. (b) Smoothed $P_{st}(M, L)$ derived from the raw data triplets in (a). $P_{st}(M, L)$ is higher in cases of longer distance and higher $M$5.5-5.8.

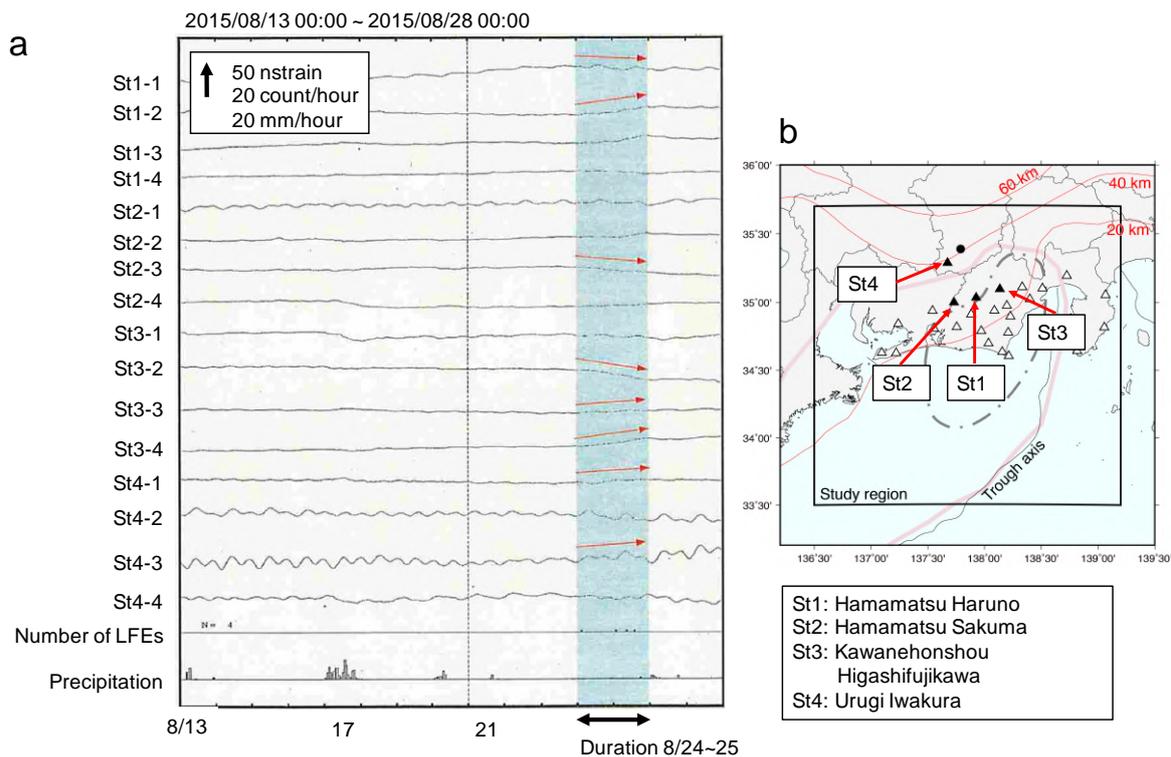

**Fig. 4** (a) Waveform in strain associated with a *M*5.5 slow slip with duration 24-25 August 2015 (the period shown in light blue). Waveform traces for the multicomponent strainmeter station "Hamamatsu Haruno" correspond to the four waveform traces from the top of the list, indicated by St1-1, St1-2, St1-3, and St1-4. The length of the upward arrow indicates 50 nanostrain (nstrain) for waveform in strain. This length also shows 20 low-frequency earthquakes (LFEs) per hour for number of LFEs, and 20 mm per hour for precipitation. (b) Location map of strainmeter stations (filled triangles) and the slow slip (circle) detected and located by these stations. Open triangles indicate strainmeter stations that were not used for detecting and locating the slow slip. See also https://www.data.jma.go.jp/svd/eqev/data/gaikyo/hantei20150831/mate01.pdf.

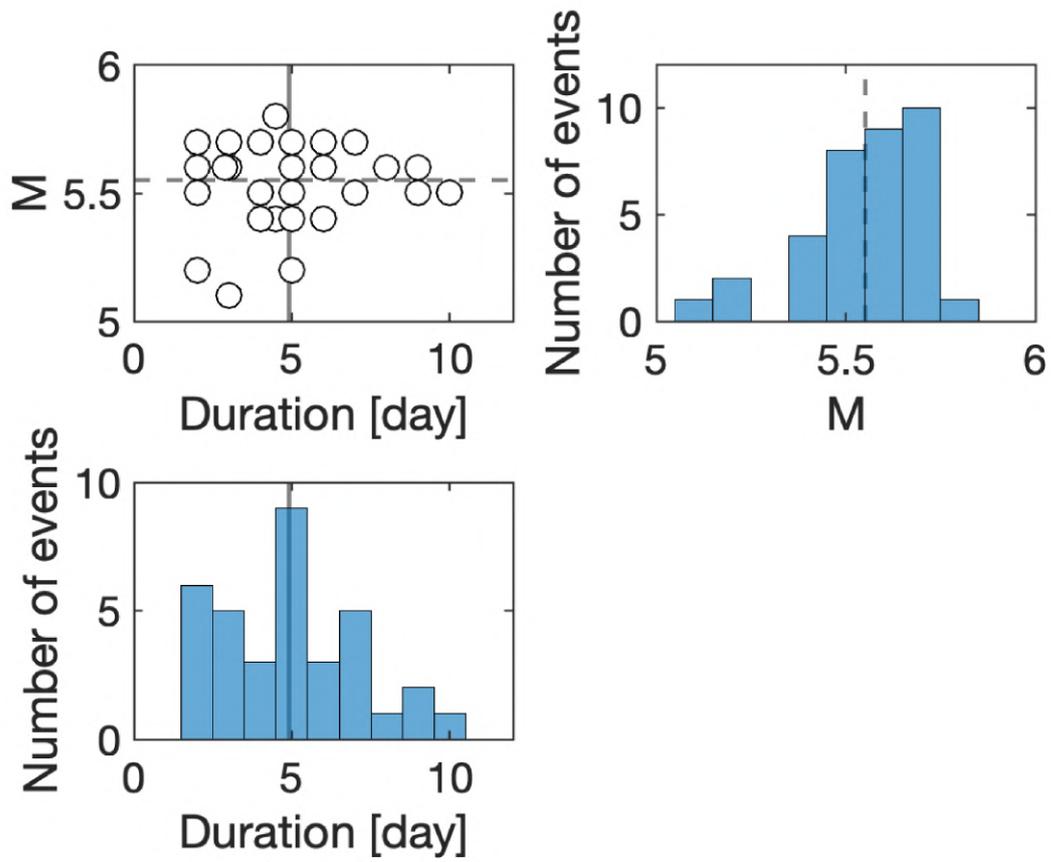

**Fig. 5** *M* versus duration of slow slips in (a). Dashed line: mean *M* of 5.55; Solid line: mean duration of 4.91 days. (b,c) Histogram of *M* and duration, respectively.

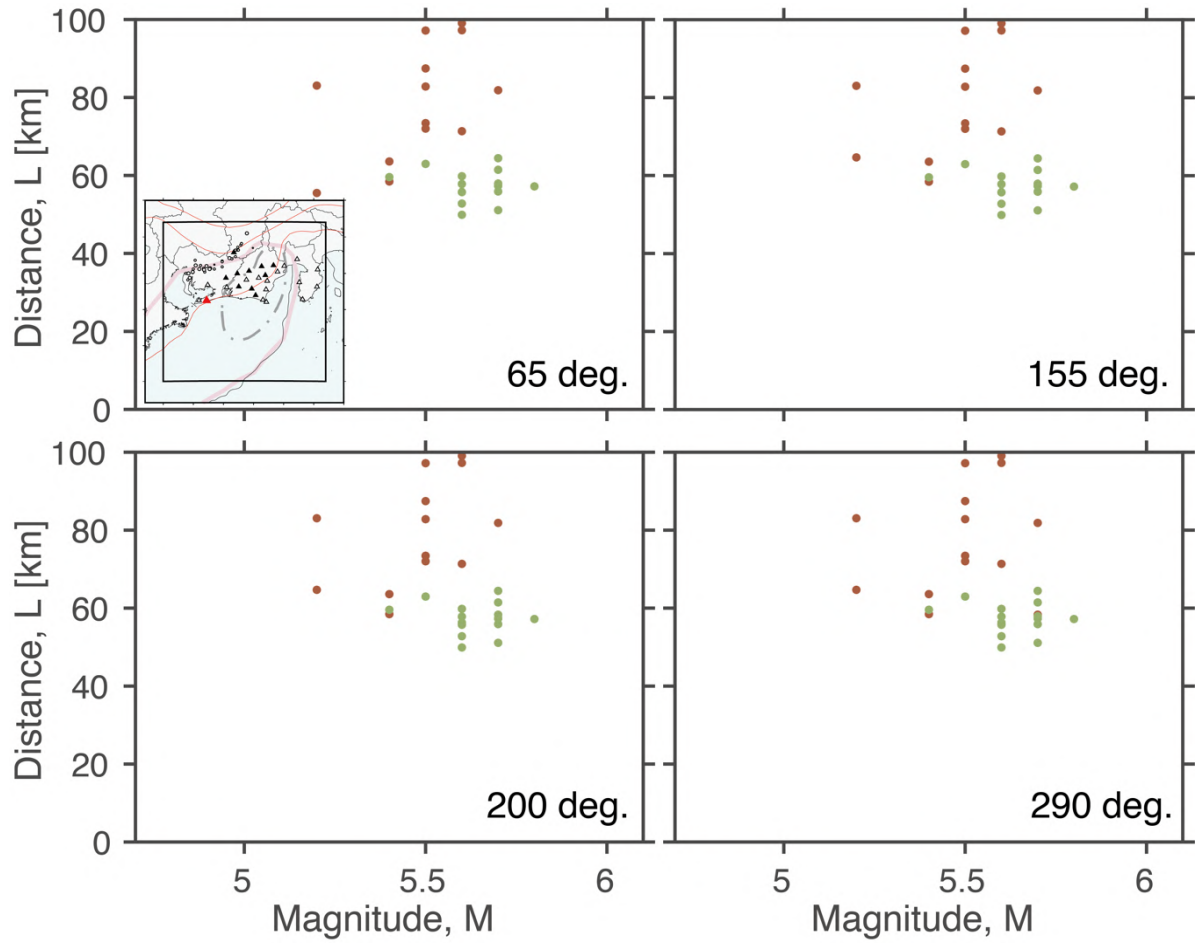

**Fig. 6** Distributions of data triplets (undetected events: red; detected events: green) for four different components with different directions (clockwise from the north) for station "Tahara Takamatsu", indicated by a red triangle in the inset. The distributions are very similar.

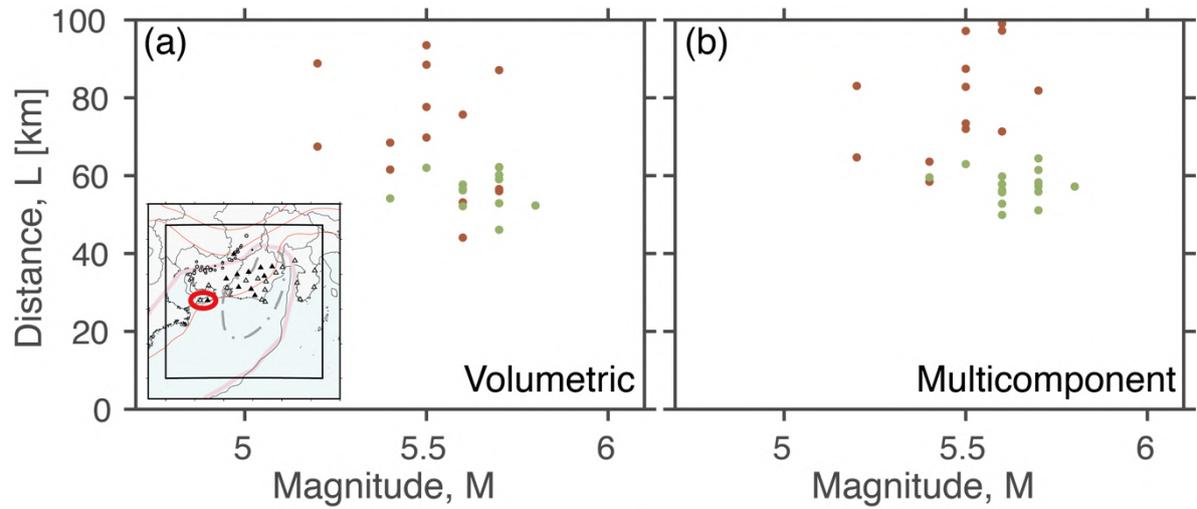

**Fig. 7** Distribution of data triplets for strainmeter stations: (a) "Tahara Fukue" (volumetric) and (b) "Tahara Takamatsu" (multicomponent). Locations of the respective stations are indicated by open and filled triangles in the red ellipsoid in the inset. For "Tahara Takamatsu", stacked data triplets over four components are shown in (b). The patterns in (a) and (b) are very similar.

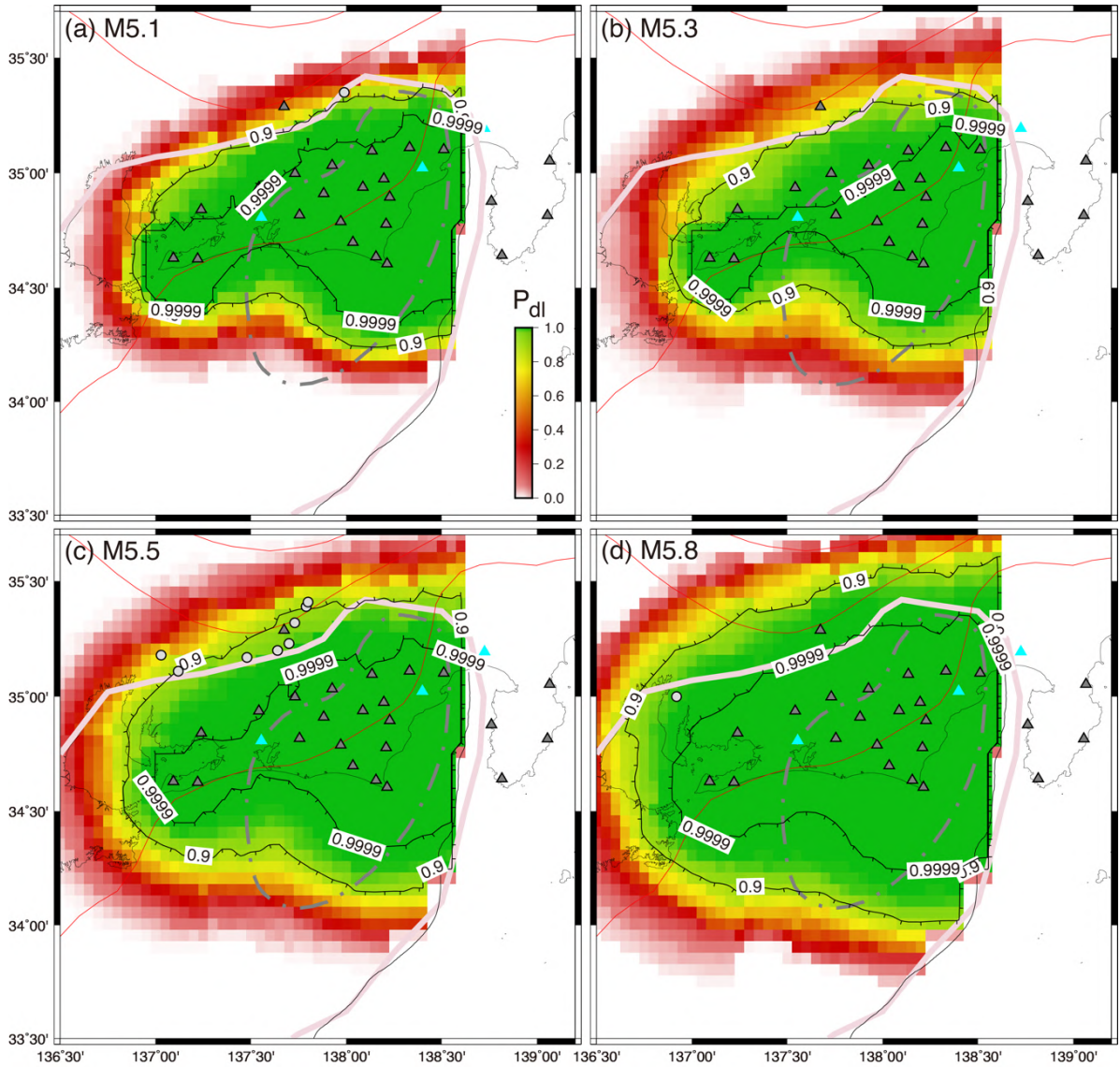

**Fig. 8** Maps of $P_{\mathrm{dl}}(M, x)$ for $M = 5.1$, 5.3, 5.5, and 5.8 in (a-d). The map is resolved on the slip interface. Triangles: stations (grey, stations used to compute $P_{\mathrm{dl}}$; cyan, stations not used). Circles: slow slips of respective $M$ that were recorded in 2012-2016. Maps for $M = 5.1$, 5.2, …, and 5.8 are given in Figure 1 of Online Resource. See also the caption of Figure 1.

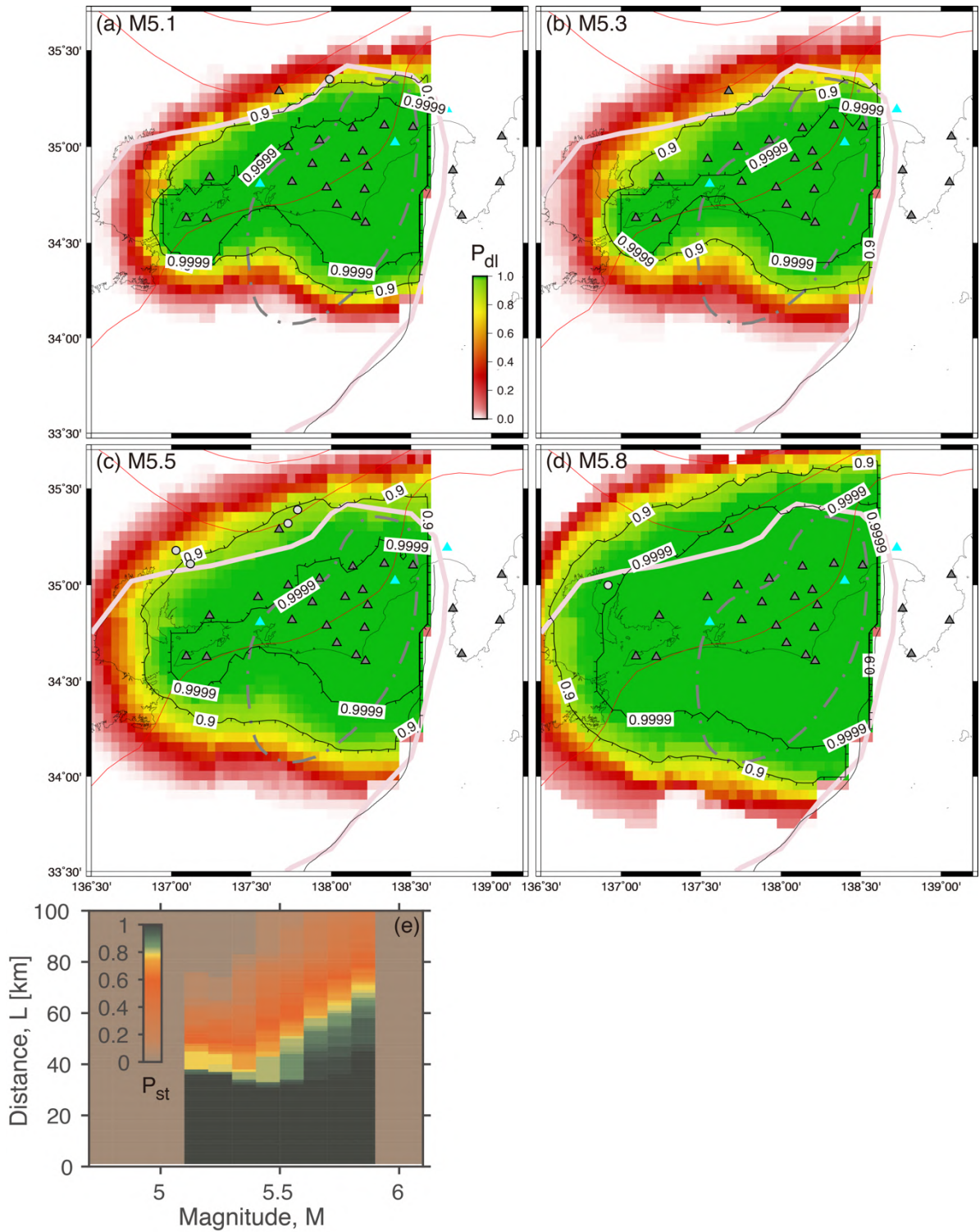

**Fig. 9** A case study that used slow slips whose duration is shorter than or equal to 5 days. (a-d) Maps of $P_{dl}(M, x)$ for $M = 5.1$, 5.3, 5.5, and 5.8. Triangles: stations (grey, stations used to compute $P_{dl}$; cyan, stations excluded from the $P_{dl}$ computation). Circles: slow slips of respective $M$ that were recorded in 2012-2016. To create the maps in (a-d), $P_{st}$ based on slow slips with duration ≤ 5 days in (e) was used. Maps for $M = 5.1$, 5.2, …, and 5.8 are given in Figure 2 of Online Resource. See also the caption of Figure 1.

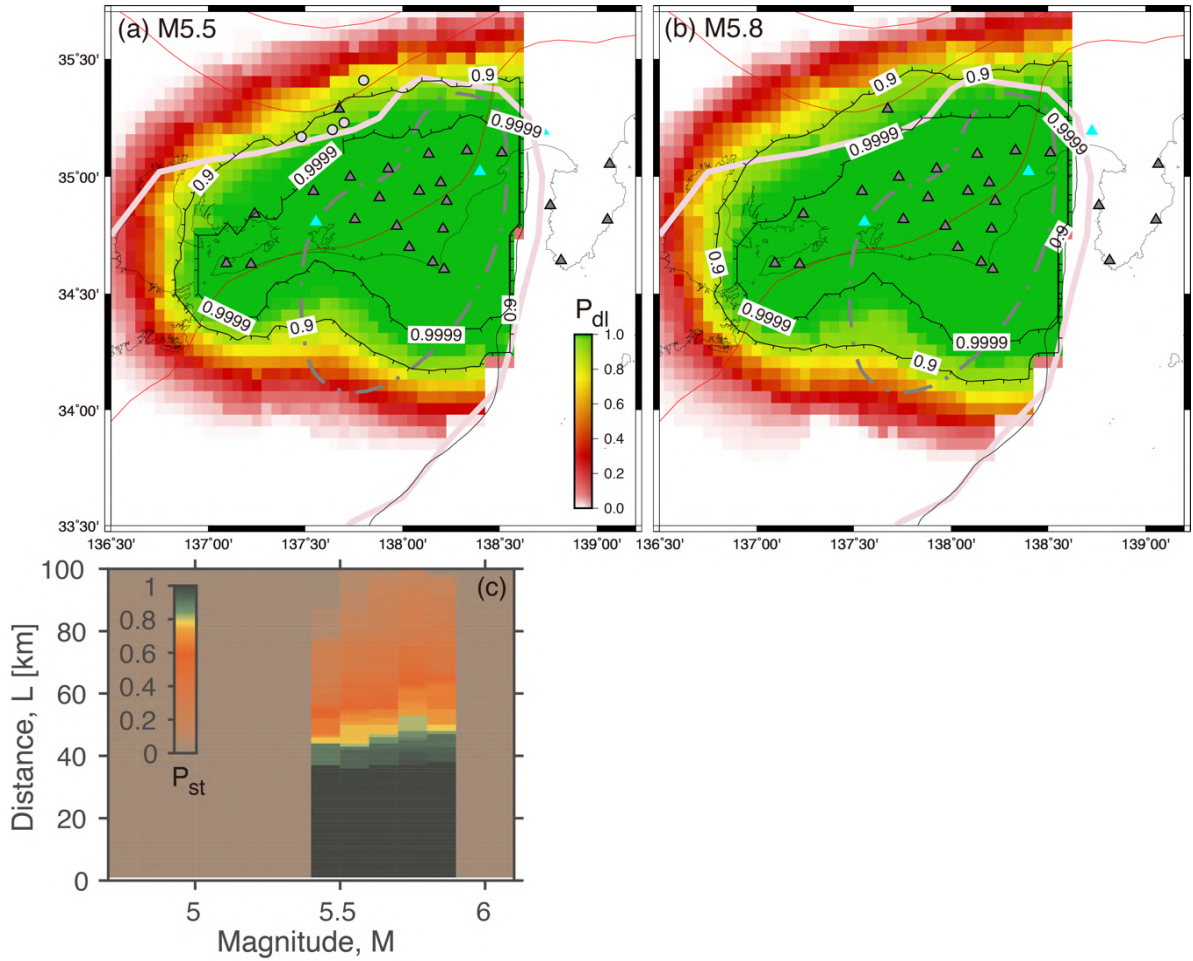

**Fig. 10** Same as Figure 9 for the case in which slow slips with duration > 5 days were used. No slow slip of $M$ = 5.1-5.3 with duration > 5 days was recorded so that $P_{st}(M < 5.4, L) = 0$ in (c). (a,b) Maps of $P_{dl}(M, x)$ for $M$ = 5.5 and 5.8. Maps for $M$ = 5.4, 5.5, …, and 5.8 are given in Figure 3 of Online Resource. See also the caption of Figure 1.

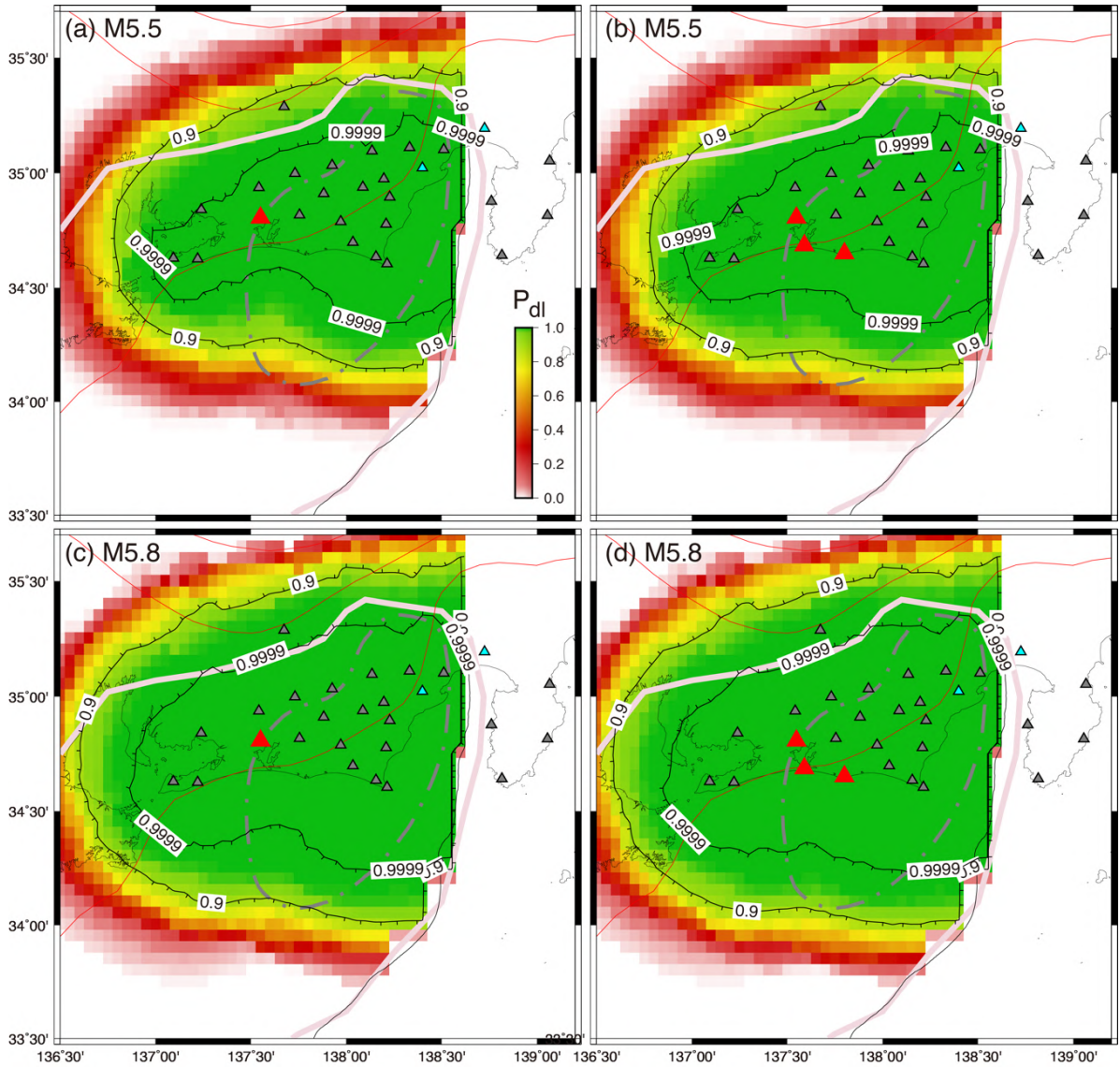

**Fig. 11** Scenario computation of adding virtual stations (red) to the network (grey). Target *M* is 5.5 in (a,b) and 5.8 in (c,d). A single virtual station was added to a location in (a,c). (b,d) Two stations were added to the configuration in (a,c), respectively. See also the caption of Figure 1.